\documentclass[12pt]{article}

\usepackage{amsfonts}
\usepackage[centertags]{amsmath}
\usepackage{amssymb}
\usepackage[verbose,colorlinks=true,naturalnames=true,linkcolor=blue,]{hyperref}

\def\ba{\begin{array}}
\def\ea{\end{array}}
\def\be{\begin{equation}}
\def\ee{\end{equation}}
\def\bea{\begin{eqnarray}}
\def\eea{\end{eqnarray}}
\usepackage[verbose,colorlinks=true,naturalnames=true,linkcolor=blue,]{hyperref}

\newcounter{rown}

\begin{document}

\title{
Quantum deformations of $D=4$ Euclidean, Lorentz, Kleinian and quaternionic $\mathfrak{o}^{\star}(4)$ symmetries\\ in unified $\mathfrak{o}(4;\mathbb{C})$ setting\\[10pt]
}
\author{A. Borowiec$^{1}$, J. Lukierski$^{1}$ and V.N. Tolstoy$^{1,2}$ \\
\\
$^{1}$Institute for Theoretical Physics, \\
University of Wroc{\l }aw, pl. Maxa Borna 9, \\
50--205 Wroc{\l }aw, Poland\\
\\
$^{2}$Lomonosov Moscow State University,\\
Skobeltsyn Institute of Nuclear Physics, \\
Moscow 119991, Russian Federation}
\date{}
\maketitle
\begin{abstract}
We employ new calculational technique and present complete list of   classical $r$-matrices for $D=4$ complex homogeneous orthogonal Lie algebra $\mathfrak{o}(4;\mathbb{C})$, the rotational symmetry of  four-dimensional
complex space-time. Further applying reality conditions we obtain the classical $r$-matrices for all possible real forms of $\mathfrak{o}(4;\mathbb{C})$: Euclidean $\mathfrak{o}(4)$, Lorentz $\mathfrak{o}(3,1)$, Kleinian $\mathfrak{o}(2,2)$ and quaternionic $\mathfrak{o}^{\star}(4)$ Lie algebras. For $\mathfrak{o}(3,1)$ we get known four classical $D=4$ Lorentz $r$-matrices, but for other real Lie algebras (Euclidean, Kleinian, quaternionic) we provide new results and mention some applications.

\end{abstract}
\setcounter{equation}{0}
\section{Introduction}
In recent years due to efforts to construct quantum gravity characterized by noncommutative space-time structures at Planckian distaces \cite{Maj88}--\cite{Gar95}, the ways in which one deforms the space-time coordinates
and space-time symmetries became important. A principal tool for the classification of quantum deformations is provided by the classical $r$-matrices \cite{BD82}--\cite{Maj95}.

In this paper we shall consider $D=4$ orthogonal Lie algebras $\mathfrak{o}(4-k,k)$ ($k=0,1,2$): for $k=0$ we obtain the $D=4$ Euclidean symmetry used in functional integration formalism and topological considerations,
for $k=1$ we get the $D=4$ Lorentz or $D=3$ $dS$ Lie algebra, and the case $k=2$ describes the symmetry of  $D=4$ space-time with neutral or Kleinian signature $(-,+,-,+)$ used in two-dimensional double field theory
(see e.g. \cite{HZ09,Ren14}) or employed as $D=3$ $AdS$ Lie algebra. The main novelty of our paper consists in the method of obtaining the classical $r$-matrices: firstly we shall study all quantum deformations ($r$-matrices)
for the $D=4$ complex Lie algebra $\mathfrak{o}(4;\mathbb{C})$ and then by imposing suitable reality conditions we shall obtain respective $r$-matrices for the real symmetries $\mathfrak{o}(4-k,k)$ ($k=0,1,2$).
For completeness we added to our considerations the Lie algebra of two-dimensional quaternionic group $\textit{O}(2;\mathbb{H}):=\textit{O}^{\star}(4)\cong\textit{O}(2,1)\oplus\textit{O}(3)$ which is the fourth real
form of $\textit{O}(4;\mathbb{C})$.

It should be pointed out that the knowledge of classical $r$-matrices describing deformations of space-time symmetries is important in present studies of gravity models and string theory, in particular for the formulation of quantum-deformed field-theoretic models and related gravity/gauge correspondence. Since introduction in 2002 two-dimensional Yang-Baxter deformed $\sigma$-models \cite{Klim02}--\cite{Vic15} there are available techniques
linking classical $r$-matrices of space--time symmetry algebras with various gravity solutions describing the string theory backgrounds \cite{KMY14}--\cite{BKLSY15}. In such framework the classical $r$-matrices are useful
in description of gravity/gauge correspondence for the gauge sector described by noncommutative gauge field theory \cite{HI99}--\cite{MYosh14}, \cite{vanTong15}.
The full list of  quantum deformations for the $D=4$ Lorentz algebra, described by the real classical $\mathfrak{o}(3,1)$ $r$-matrices, is known already since 20 years (\cite{Zak94}; see also \cite{Tol07}).
However, an analogous classification of real classical $r$-matrices (quantum deformations) for $D=4$ Euclidean and Kleinian\footnote{We point out that $D=4$ Kleinian rotations $\mathfrak{o}(2,2)$ describe as well $D=3$ $AdS$
algebra and $D=2$ conformal algebra.} signatures as well as for the algebra $\mathfrak{o}^{\star}(4)\simeq\mathfrak{o}(2|\mathbb{H})$ (see e.g. \cite{Gilm74}) have not been given.

Our new method consists in two steps:
\\
(i) Construct the complete list of complex $\mathfrak{o}(4;\mathbb{C})$ $r$-matrices.
\\
(ii) Impose respective  four reality conditions to get the classification of real r-matrices for $\mathfrak{o}(4-k,k)$ algebras (k=0,1,2) and for $\mathfrak{o}^{\star}(4)$.

In present paper we perform in detail these two steps and present the results in explicite form. It follows from our construction which classical $r$-matrices are satisfying standard (homogeneous) CYBE, and which are satisfying modified (nonhomogeneous) YBE.

The plan of our paper is the following: In Sect.2 we consider complex Lie algebra $\mathfrak{o}(4;\mathbb{C})$ and its four real forms. In Sect.~3 we provide, after the use of possible automorphisms, the complete list of complex $\mathfrak{o}(4;\mathbb{C})$ $r$-matrices. In Sect.~4 we consider the real forms. For Lorentz signature we shall confirm the old results of Zakrzewski \cite{Zak94}, for Euclidean signature we obtain one $r$-matrix with three independent parameters, for Kleinian case we have six $r$-matrices (four with 3, two with 2 and one with one parameters) and finally for the case $\mathfrak{o}^{\star}(4)$ there are two $r$-matrices both with 3 parameters. It should be noted that the $r$-matrices for $\mathfrak{o}^{\star}(4)$ have been never considered. In Sect.~5 we present our plans how to extend the presented results.

\setcounter{equation}{0}
\section{Complex $D=4$ Euclidian algebra and its real forms}
In this section we describe $D=4$ complex Euclidean Lie algebra and its real forms: Euclidian, Lorentz, Kleinian and quaternionic orthogonal Lie algebras\footnote{It should be noted that the names ''Euclidean'' and Kleinian
are used also for denomination of inhomogeneous Lie symmetries: rotations with translations generated by fourmomenta.} in terms of different bases. {\it The complex $D=4$ Euclidean Lie algebra} $\mathfrak{o}(4;\mathbb{C})$ is
generated by six Euclidean basis elements $L_{\mu\nu}=-L_{\nu\mu}\in\mathfrak{o}(4;\mathbb{C})$ ($\mu,\nu=0,1,2,3$)
satisfying the relations:
\begin{eqnarray}\label{pr1}
\begin{array}{rcl}
[L_{\mu\nu},\,L_{\lambda\rho}]\!\!& =\!\!&g_{\nu\lambda}\,L_{\mu\rho}-g_{\nu\rho}\,L_{\mu\lambda}+g_{\mu\rho}\,L_{\nu\lambda}-g_{\mu\lambda}\,L_{\nu\rho}~,
\end{array}%
\end{eqnarray}
where  $g_{\mu\nu}$ is the Euclidean metric: $g_{\mu\nu}=\mathop{\rm diag}\,(1,1,1,1)$. The Euclidean algebra $\mathfrak{o}(4;\mathbb{C})$, as a linear space, is a linear envelope of the basis
$\{L_{\mu\nu}\}$ over $\mathbb{C}$. By analogy with the Lorentz algebra it is convenient to introduce the following generators
\begin{eqnarray}\label{pr2}
M_{i}\!\!&:=\!\!&-\frac{1}{2}\varepsilon_{ijk}L_{jk}~,\qquad N_{i}\;:=\;-L_{0i}\qquad(i,j,k=1,2,3)~.
\end{eqnarray}
In this notation, the defining relations (\ref{pr1}) take the form:
\begin{eqnarray}\label{pr3}
\begin{array}{ccccccc}
[M_{i},\,M_{j}]\!\!&=\!\!&\varepsilon_{ijk}M_{k}~,\qquad[M_{i},\,N_{j}]\!\!&=\!\!&\varepsilon_{ijk}N_{k}~,\qquad[N_{i},\,N_{j}]\!\!&=\!\!&\varepsilon_{ijk}M_{k}~.
\end{array}%
\end{eqnarray}
The set $\{M_{i},N_{i}|i=1,2,3\}$ will be called the {\it Cartesian} basis
\footnote{In the case of the Lorentz algebra $\mathfrak{o}(3,1)$ the elements $\{M_{i}\}$ are the infinitesimal rotations in the space coordinate planes $(x_{i},x_{j})$, $1<i<j$, and the elements $\{N_{i}\}$ (
boosts) are related with rotations in the time-space coordinate planes $(x_{0},x_{i})$, $i=1,2,3$.}.

If we consider a Lie algebra with the commutation relations (\ref{pr3}) over $\mathbb{R}$ then we have the compact real form $\mathfrak{o}(4):=\mathfrak{o}(4;\mathbb{R})\cong\mathfrak{o}(3)\oplus\mathfrak{o}(3)$ with the anti-Hermitian basis
\begin{eqnarray}
&&\begin{array}{ccccc}\label{pr4}
M^{*}_{i}\!\!&=\!\!&-M_{i}~,\qquad N^{*}_{i}\!\!&=\!\!&-N_{i}\qquad\;(i=1,2,3)
\end{array}\qquad\quad{\rm{for}}\;\;\mathfrak{o}(4)~.
\end{eqnarray}
It is well-known from theory of real forms for semisimple complex Lie algebras (including our case $\mathfrak{o}(4;\mathbb{C})$) that all real (non-compact) forms can be constructed by involutive automorphisms of the
compact form. For each such Lie algebra the compact form is unique (see e.g. \cite{Gant39}). In our case there are only three real non-compact forms (see e.g. \cite{BarutRacz77}):
{\it the Lorentz algebra $\mathfrak{o}(3,1):=\mathfrak{o}(3,1;\mathbb{R})\cong\mathfrak{sl}(2;\mathbb{C})^{\mathbb{R}}$}, {\it the Kleinian algebra $\mathfrak{o}(2,2):=\mathfrak{o}(2,2;\mathbb{R})\cong\mathfrak{o}(2,1)\oplus\mathfrak{o}(2,1)$} and {\it the quaternionic Lie algebra $\mathfrak{o}^{\star}(4):=\mathfrak{o}(2;\mathbb{H})\cong\mathfrak{o}(2,1)\oplus\mathfrak{o}(3)$}. We remind that $\mathfrak{o}(3)\cong\mathfrak{su}(2)$, $\mathfrak{o}(2,1)\cong\mathfrak{sl}(2;\mathbb{R})\cong\mathfrak{su}(1,1)\cong\mathfrak{sp}(2;\mathbb{R})$. The involutive automorphisms, which introduce these non-compact forms, are defined as follows ($i=1,2,3;$ $k=1,3$):
\begin{eqnarray}\label{pr5}
&&\omega_{\dag}^{}(M_{i})\;=\;M_{i}~,\qquad\qquad\;\omega_{\dag}^{}(N_{i})\;=\;-N_{i}\qquad\qquad\;\;{\rm{for}}\;\;\mathfrak{o}(3,1)~,
\\[7pt]
&&\omega_{\ddag}^{}(M_{i})\;=\;(-1)^{i}M_{i}~,\quad\omega_{\ddag}^{}(N_{i})\;=\;(-1)^{i}N_{i}\qquad{\rm{for}}\;\;\mathfrak{o}(2,2)~,
\\[7pt]
&&\begin{array}{ccccc}\label{pr7}
\omega_{\star}^{}(M_{2})\!\!&=\!\!&M_{2}~,\qquad\quad\;\;\omega_{\star}^{}(M_{k})\!\!&=\!\!&-N_{k}~,\\[5pt]
\omega_{\star}^{}(N_{2})\!\!&=\!\!&N_{2}~,\qquad\quad\;\;\omega_{\star}^{}(N_{k})\!\!&=\!\!&-M_{k}
\end{array}\qquad\quad\;\;{\rm{for}}\;\;\mathfrak{o}^{\star}(4)~.
\end{eqnarray}
The corresponding conjugations (antilinear, involutive antiautomorphisms) $\dag,\ddag,\star$ are connected with the involutive automorphisms $\omega_{\sharp}^{}$ ($\sharp=\dag,\ddag,\star$) by the formulas:
\begin{eqnarray}\label{pr8}
(\cdot)^{\dag}\!\!&=\!\!&\omega_{\dag}^{}((\cdot)^{*})~,\qquad(\cdot)^{\ddag}\,=\,\omega_{\ddag}^{}((\cdot)^{*})~,\qquad(\cdot)^{\star}\,=\,\omega_{\star}^{}((\cdot)^{*})~.
\end{eqnarray}
Another convenient basis for the complex Euclidean algebra $\mathfrak{o}(4;\mathbb{C})$ is called {\it chiral} (left $X_{i}$ and right $\bar{X}_{i}$) $\mathfrak{sl}(2)\equiv\mathfrak{sl}(2;\mathbb{C})$ basis, defined as follows:
\begin{eqnarray}\label{pr9}
X_{i}\!\!&:=\!\!&\frac{1}{2}(M_{i}+N_{i})~,\qquad \bar{X}_{i}\,=\,\frac{1}{2}(M_{i}-N_{i})\qquad(i=1,2,3)~.
\end{eqnarray}
In this basis the defining relations (\ref{pr3}) look as follows:
\begin{eqnarray}\label{pr10}
\begin{array}{rcl}
[X_{i},\,X_{j}]\!\!&=\!\!&\varepsilon_{ijk}X_{k}~,\qquad[\bar{X}_{i},\,\bar{X}_{j}]\,=\,\varepsilon_{ijk}\bar{X}_{k}~,\qquad[X_{i},\,\bar{X}_{j}]\,=\,0~.
\end{array}%
\end{eqnarray}
The chiral basis decomposes the complex Lie algebra $\mathfrak{o}(4;\mathbb{C})$ into a direct sum: $\mathfrak{o}(4;\mathbb{C})=\mathfrak{sl}(2)\oplus\bar{\mathfrak{sl}}(2)$. The real forms $\mathfrak{o}(4)$, $\mathfrak{o}(3,1)$, $\mathfrak{o}(2,2)$, $\mathfrak{o}^{\star}(4)$ in this basis easy are given by the formulas (\ref{pr4})--(\ref{pr8}) and they are as follows ($i=1,2,3$):
\begin{eqnarray}\label{pr11}
X^{*}_{i}\!\!&=\!\!&-X_{i}^{}~,\qquad\qquad\quad\;\;\bar{X}^{*}_{i}\;=\;-\bar{X}_{i}^{} \qquad\qquad\quad{\rm{for}}\;\;\mathfrak{o}(4)~,
\\[5pt]\label{pr12}
X_{i}^{\dag}\!\!&=\!\!&-\bar{X}_{i}^{}~,\qquad\qquad\quad\;\;\bar{X}_{i}^{\dag}\;=\;-X_{i}^{}\qquad\qquad\quad{\rm{for}}\;\;\mathfrak{o}(3,1)~,
\\[5pt]\label{pr13}
X_{i}^{\ddag}\!\!&=\!\!&(-1)^{i-1}X_{i}^{}~,\qquad\quad\bar{X}_{i}^{\ddag}\;=\;(-1)^{i-1}\bar{X}_{i}^{}\qquad\;\;{\rm{for}}\;\;\mathfrak{o}(2,2)~,
\\[5pt]\label{pr14}
X_{i}^{\star}\!\!&=\!\!&(-1)^{i-1}X_{i}^{}~,\qquad\quad\bar{X}_{i}^{\star}\;=\;-\bar{X}_{i}^{}\qquad\qquad\quad\;{\rm{for}}\;\;\mathfrak{o}^{\star}(4)~.
\end{eqnarray}
For description of quantum deformations and in particular for the classification of classical $r$-matrices of the  complex Euclidean algebra $\mathfrak{o}(4;\mathbb{C})$ and its real forms it is convenient to use
the Cartan--Weyl bases in both sectors of the sum $\mathfrak{o}(4;\mathbb{C})=\mathfrak{sl}(2)\oplus\bar{\mathfrak{sl}}(2)$.  Such basis is given by
\begin{eqnarray}\label{pr15}
H\!\!&:=\!\!&-\imath X_{3}~,\quad E_{\pm}\,=\,-\imath X_{1}\mp X_{2}~,\quad\bar{H}\,:=\,\imath\bar{X}_{3}~,\quad\bar{E}_{\pm}\,=\,\imath\bar{X}_{1}\mp\bar{X}_{2}
\end{eqnarray}
with the non-zero defining relations:
\begin{eqnarray}\label{pr16}
[H,\,E_{\pm}]\!\!&=\!\!&E_{\pm}~,\quad[E_{+},\,E_{-}]\,=\,2H~,\quad[\bar{H},\,\bar{E}_{\pm}]\,=\,\bar{E}_{\pm}~,\quad [\bar{E}_{+},\,\bar{E}_{-}]\,=\,2\bar{H}~.
\end{eqnarray}
In the basis (\ref{pr15}), (\ref{pr16}) all possible real forms of $\mathfrak{o}(4;\mathbb{C})$ satisfy the following reality conditions:
\begin{eqnarray}\label{pr17}
H^{*}\!\!&=\!\!&H,\qquad E_{\pm}^{*}\;=\;E_{\mp},\qquad\;\bar{H}^{*}\;=\;\bar{H},\qquad\bar{E}_{\pm}^{*}\;=\;\bar{E}_{\mp}\qquad{\rm{for}}\;\;\mathfrak{o}(4),
\\[5pt]\label{pr18}
H^{\dag}\!\!&=\!\!&-\bar{H},\quad\;E_{\pm}^{\dag}\;=\;-\bar{E}_{\pm},\quad\;
\bar{H}^{\dag}\;=\;-H,\quad\;\bar{E}_{\pm}^{\dag}\;=\;-E_{\pm}\quad\;{\rm{for}}\;\;\mathfrak{o}(3,1),
\\[5pt]\label{pr19}
H^{\ddag}\!\!&=\!\!&-H,\quad\;E_{\pm}^{\ddag}\;=\;-E_{\pm},\quad\;\;\bar{H}^{\ddag}\;=\;-\bar{H},\quad\bar{E}_{\pm}^{\ddag}\;=\;-\bar{E}_{\pm}\quad\;{\rm{for}}\;\;\mathfrak{o}(2,2),
\\[5pt]\label{pr20}
H^{\star}\!\!&=\!\!&-H,\quad\;E_{\pm}^{\star}\;=\;-E_{\pm},\quad\;\;\bar{H}^{\star}\;=\;\bar{H},\qquad\bar{E}_{\pm}^{\star}\;=\;\bar{E}_{\mp}\qquad{\rm{for}}\;\;\mathfrak{o}^{\star}(4),
\end{eqnarray}
where the conjugations $*,\dag,\ddag,\star$  are determined by the equations (\ref{pr11})--(\ref{pr15}).

\setcounter{equation}{0}
\section{Classical $r$-matrices of $\mathfrak{o}(4;\mathbb{C})$}
By the definition each classical $r$-matrix of the complex $D=4$ Euclidean Lie algebra $\mathfrak{o}(4;\mathbb{C})$, $r\in\mathfrak{o}(4;\mathbb{C})\wedge\mathfrak{o}(4;\mathbb{C})$,
satisfy the classical Yang-Baxter equation (YBE):
\begin{eqnarray}\label{cr1}
[[r,\,r]]\!\!&=\!\!&\Omega~.
\end{eqnarray}
Here $[[\cdot,\cdot]]$ is the Schouten bracket which for any monomial skew-symmetric two-tensors $r_{1}^{}=x\wedge y$ and $r_{2}^{}=u\wedge v$ ($x,y,u,v\in\mathfrak{o}(4;\mathbb{C})$) is given by\footnote{For general polynomial (a sum of monomials) two-tensors $r_{1}$ and $r_{2}$ one can use the bilinearity of the Schouten bracket.}
\begin{eqnarray}\label{cr2}
\begin{array}{rcl}
[[x\wedge y,\,u\wedge v]]\!\!&:=\!\!&x\wedge\bigl([y,u]\wedge{v}+u\wedge[y,v]\bigr)
\\[7pt]
&&-y\wedge\bigl([x,u]\wedge{v}+u\wedge[x,v]\bigr)\,=\,[[u\wedge v,\,x\wedge y]]
\end{array}
\end{eqnarray}
and $\Omega$ is the $\mathfrak{o}(4;\mathbb{C})$-invariant element, $\Omega\in(\stackrel{3}\wedge\mathfrak{o}(4;\mathbb{C}))_{\mathfrak{o}(4;\mathbb{C})}$\footnote{It means that $[\Delta^2(x),\,\Omega]=0$ for $\forall x\in\mathfrak{o}(4;\mathbb{C})$, where $\Delta(\cdot)$ is the primitive coproduct in $\mathfrak{o}(4;\mathbb{C})$.}:
\begin{eqnarray}\label{cr3}
\Omega\!\!&=\!\!&{\gamma}\,\Omega(\mathfrak{sl}(2))+\bar{\gamma}\,\Omega(\bar{\mathfrak{sl}}(2))\,=\,\gamma\,{E}_{+}\wedge H\wedge E_{-}+\bar{\gamma}\,\bar{E}_{+}\wedge\bar{H}\wedge\bar{E}_{-}
\end{eqnarray}
in the basis  (\ref{pr16}).

Since the Lie algebra $\mathfrak{o}(4;\mathbb{C})$ is the direct sum, $\mathfrak{o}(4;\mathbb{C})=\mathfrak{sl}(2)\oplus\bar{\mathfrak{sl}}(2)$, and $\mathfrak{o}(4;\mathbb{C})\wedge\mathfrak{o}(4;\mathbb{C})=\mathfrak{sl}(2)\wedge\mathfrak{sl}(2)\oplus\bar{\mathfrak{sl}}(2)\wedge\bar{\mathfrak{sl}}(2)\oplus\mathfrak{sl}(2)\wedge\bar{\mathfrak{sl}}(2)$ therefore the $r$-matrix $r$ has the following decomposition:
\begin{eqnarray}\label{cr4}
r\!\!&=\!\!&a+\bar{a}+b~,
\end{eqnarray}
where $a\in A:=\mathfrak{sl}(2)\wedge\mathfrak{sl}(2)$, $\bar{a}\in\bar{A}:=\bar{\mathfrak{sl}}(2)\wedge\bar{\mathfrak{sl}}(2)$, $b\in B:=\mathfrak{sl}(2)\wedge\bar{\mathfrak{sl}}(2)$. Substituting the decomposition (\ref{cr4}) in the bilinear equation (\ref{cr1}) and collecting homogeneous terms we obtain the system of equations:
\begin{eqnarray}\label{cr5}
[[a,\,a]]\!\!&=\!\!&\,\gamma\,E_{+}\wedge H\wedge E_{-}~,
\\[5pt]\label{cr6}
[[\bar{a},\,\bar{a}]]\!\!&=\!\!&\,\bar{\gamma}\,\bar{E}_{+}\wedge\bar{H}\wedge\bar{E}_{-}~,
\\[5pt]\label{cr7}
[[b,\,b]]\!\!&=\!\!&-2[[a,\,b]]-2[[\bar{a},\,b]]~.
\end{eqnarray}
From the first two equations (\ref{cr5}) and (\ref{cr6}) we see that the components $a$ and $\bar{a}$ of the classical $r$-matrix (\ref{cr4}) are the classical $r$-matrices and therefore in order to get the total list of the classical $r$-matrices of $\mathfrak{o}(4;\mathbb{C})$ we need first to take all solutions of the equations (\ref{cr5}), (\ref{cr6}) and then to solve the consistency conditions (\ref{cr7}).

{\it Sectors $A$, $\bar{A}$, $A\oplus\bar{A}$}. Let us write down all classical $r$-matrices of the sectors $A$, $\bar{A}$ and $A\oplus\bar{A}$. It is well-known that up to isomorphism there are only two classical $r$-matrices for $\mathfrak{sl}(2)$: the standard one and the Jordanian type (see e.g.(\cite{D-VL90}) and therefore for the sectors $A$, $\bar{A}$ we get\\
\begin{eqnarray}\label{cr8}
a_{0}^{}\!\!&=\!\!&\gamma\,E_{+}\wedge E_{-}~,\qquad a_{+}^{}\;=\;\chi\,E_{+}\wedge H\qquad({\rm for}\;\;Sector\;A)~,
\end{eqnarray}
\begin{eqnarray}\label{cr9}
\bar{a}_{0}^{}\!\!&=\!\!&\bar{\gamma}\,\bar{E}_{+}\wedge\bar{E}_{-}~,\qquad\bar{a}_{+}^{}\;=\;\bar{\chi}\,\bar{E}_{+}\wedge\bar{H}\qquad({\rm for}\;\;Sector\;\bar{A})~.
\end{eqnarray}
Here the two-tensors $a_{0}^{}$ and $\bar{a}_{0}^{}$ are the standard $r$-matrices which satisfy the non-homogeneous YBEs (\ref{cr5}) and (\ref{cr6}) with $\gamma,\bar{\gamma}\neq0$, and the two-tensors $a_{+}^{}$ and $\bar{a}_{+}^{}$ are the $r$-matrices of the Jordanian type, which satisfy the homogeneous YBEs (\ref{cr5}) and (\ref{cr6}) with $\gamma,\bar{\gamma}=0$. In general case the parameters $\gamma$, $\bar{\gamma}$ and $\chi$, $\bar{\chi}$ are complex numbers  however parameter $\chi$ (analogously $\bar{\chi}$) can be removed by the ''rescaling'' isomorphism: $\varphi(E_{+})=\chi^{-1}E_{+}$, $\varphi(E_{-})=\chi {E}_{-}$, $\varphi(H)=H$. It should be noted also that any linear combination of the standard and Jordanian $r$-matrices, $a_{0}$ and $a_{+}$ (analogously for $\bar{a}_{0}$ and $\bar{a}_{+}$) is also a classical $r$-matrix and it can be reduced to $a_{0}$ ($\bar{a}_{0}$) by a $\mathfrak{sl}(2)$-automorphism. Indeed, it is easy to see that the linear mapping: $\varphi(E_{+})=E_{+}$, $\varphi(E_{-})=E_{-}-\chi^{2}E_{+}+2\chi{H}$, $\varphi(H)=H-\chi{E}_{+}$, is the isomorphism, that is $[\varphi(E_{+}),\,\varphi(E_{-})]=2\varphi(H)$, $[\varphi(H),\,\varphi(E_{\pm})]=\pm\varphi(E_{\pm})$, and we have $\varphi(E_{+})\wedge\varphi(E_{-})=E_{+}\wedge E_{-}+\chi{E}_{+}\wedge H$.

Since $[[a,\bar{a}]]=0$ therefore a sum of the classical $r$-matrices from the sectors $A$ and $\bar{A}$ is also a classical $r$-matrix and we have the following four classical $r$-matrices in Sector $A\oplus\bar{A}$: \begin{eqnarray}\label{cr10}
a_{0}^{}+\,\bar{a}_{0}^{},\qquad\,a_{0}^{}+\,\bar{a}_{+}^{},\qquad\,a_{+}^{}+\,\bar{a}_{0}^{},\qquad\,a_{+}^{}+\,\bar{a}_{+}^{} .
\end{eqnarray}

{\it Sector $B$}. Now we consider the sector $B:=\mathfrak{sl}(2)\wedge\bar{\mathfrak{sl}}(2)$. We analyze two cases: {\it (i)} when $[[b,b]]=0$ and {\it (ii)}  if $[[b,b]]\neq0$. \\
{\it (i) Case $[[b,b]]=0$}. Let $[[b,b]]=0$, namely, $b$ is a classical triangular $r$-matrix\footnote{A classical $r$-matrix satisfying the homogeneous YBE is called triangular.}, then  $[[a,b]]=0$, $[[\bar{a},b]]=0$, simultaneously. It is easy to see that arbitrary element of the sector $B$ has the form
\begin{eqnarray}\label{cr11}
\begin{array}{rcl}
b\!\!&=\!\!&\beta_{+}E_{+}\wedge(\bar{\beta}_{+}\bar{E}_{+}+\bar{\beta}_{0}\bar{H}+\bar{\beta}_{-}\bar{E}_{-})
\\[5pt]
\!\!&&\!\!+\beta_{0}H\wedge(\bar{\beta}_{+}'\bar{E}_{+}+\bar{\beta}_{0}'\bar{H}+\bar{\beta}_{-}'\bar{E}_{-})
\\[5pt]
\!\!&&\!\!+\beta_{-}E_{-}\wedge(\bar{\beta}_{+}''\bar{E}_{+}+\bar{\beta}_{0}''\bar{H}+\bar{\beta}_{-}''\bar{E}_{-})~.
\end{array}
\end{eqnarray}
It should be noted that each of three terms on the right side in (\ref{cr11}) is a classical $r$-matrix.

We substitute the expression (\ref{cr11}) in the classical Yang-Baxter equation $[[b,b]]=0$. As a result we obtained that the expression  (\ref{cr11}) is a classical $r$-matrix
if and only if it has the form
\begin{eqnarray}\label{cr12}
b\!\!&=\!\!&(\beta_{+}E_{+}+\beta_{0}H+\beta_{-}E_{-})\wedge(\bar{\beta}_{+}\bar{E}_{+}+\bar{\beta}_{0}\bar{H}+\bar{\beta}_{-}\bar{E}_{-})~.
\end{eqnarray}

Each components of this two-tensor, for instance the first component $(\beta_{+}E_{+}+\beta_{0}H+\beta_{-}E_{-})$  can be reduced to the generator $H$ or $E_{+}$ by a $\mathfrak{sl}(2)$-automorphism. Indeed, let us consider the case when $D:=\beta_{0}^{2}+4\beta_{+}\beta_{-}\neq0$. We set $\varphi(H)=D^{-\frac{1}{2}}(\beta_{+}E_{+}+\beta_{0}H+\beta_{-}E_{-})$, $\varphi(E_{+})=\beta_{+}'E_{+}+\beta_{0}'H+\beta_{-}'E_{-}$, $\varphi(E_{-})=\beta_{+}''E_{+}+\beta_{0}''H+\beta_{-}''E_{-}$. Substituting this ansatz in the system of equations
\begin{eqnarray}\label{cr13}
[\varphi(H),\,\varphi(E_{\pm})]=\pm\varphi(E_{\pm}),\qquad[\varphi(E_{+}),\,\varphi(E_{-})]=2\varphi(H),
\end{eqnarray}
we find the coefficients $(\beta')$'s and $(\beta'')$'s. The final result is given by the formulas:
\begin{eqnarray}\label{cr14}
\begin{array}{rcl}
\varphi(H)\!\!&=\!\!&\displaystyle\frac{1}{\sqrt{D}}\Bigl(\beta_{+}E_{+}+\beta_{0}H+\beta_{-}E_{-}\Bigr),
\\[12pt]
\varphi(E_{+})\!\!&=\!\!&\displaystyle\frac{\chi}{\sqrt{D}}\left(\frac{\beta_{0}+\sqrt{D}}{2}\,E_{+}-2\beta_{-}H-\frac{2\beta_{-}^{2}}{\beta_{0}+\sqrt{D}}\,E_{-}\right),
\\[12pt]
\varphi(E_{-})\!\!&=\!\!&\displaystyle\frac{1}{\chi\sqrt{D}}\left(\frac{-2\beta_{+}^{2}}{\beta_{0}+\sqrt{D}}\,E_{+}-2\beta_{+}H+\frac{\beta_{0}+\sqrt{D}}{2}E_{-}\right),
\end{array}
\end{eqnarray}
where $\chi$ is a non-zero rescaling parameter, $D:=\!\beta_{0}^{2}\!+4\beta_{+}\beta_{-}\!\neq\!0$ and if $\beta_{+}\beta_{-}\!=\!0$ then $\sqrt{D}=\beta_{0}$.

Now we consider the case when $\beta_{0}^{2}+4\beta_{+}\beta_{-}=0$ in the first component of the left side of (\ref{cr12}). In this case we set the following ansatz: $\varphi(E_{+})=\chi(\beta_{+}E_{+}+\beta_{0}\,H+\beta_{-}E_{-})$, $\varphi(H)=\beta_{+}'E_{+}+\beta_{0}'H+\beta_{-}'E_{-}$, $\varphi(E_{-})=\beta_{+}''E_{+}+\beta_{0}''H+\beta_{-}''E_{-}$. Again, substituting these expressions in the system (\ref{cr13}) we find that the desired automorphism is given as follows
\begin{eqnarray}\label{cr15}
\begin{array}{rcl}
\varphi(E_{+})\!\!&=\!\!&\displaystyle\frac{1}{\kappa\beta_{+}-\beta_{-}}\left(\beta_{+}E_{+}+\beta_{0}\,H+\beta_{-}E_{-}\right)~,
\\[12pt]
\varphi(E_{-})\!\!&=\!\!&\displaystyle\frac{1}{\kappa\beta_{+}-\beta_{-}}\left(\beta_{-}E_{+}+\kappa\beta_{0}\,H+\beta_{+}E_{-}\right),
\\[12pt]
\varphi(H)\!\!&=\!\!&\displaystyle\frac{1}{\kappa\beta_{+}-\beta_{-}}\left(-\frac{\beta_{0}}{2}\,E_{+}+(\kappa\beta_{+}+\beta_{-})\,H-\frac{\kappa\beta_{0}}{2}\,E_{-}\right),
\end{array}
\end{eqnarray}
where $\kappa$ equals to $+1$ or $-1$, and if $\beta_{+}=\pm\beta_{-}$ we take $\kappa=\mp1$ accordingly in order to avoid singularities in these formulas.

Hence we obtain that the general six-parameter classical $r$-matrix (\ref{cr12}) by $\mathfrak{sl}(2)$-automor\-phims is reduced to four independent classical $r$-matrices:
\begin{eqnarray}\label{cr16}
&b_{++}^{}=\,\chi' E_{+}\wedge\bar{E}_{+},\quad b_{+0}^{}=\,\chi'' E_{+}\wedge\bar{H},\quad b_{0+}^{}=\,\bar{\chi}''H\wedge\bar{E}_{+},\quad b_{00}^{}=\,\eta H\wedge\bar{H},
\end{eqnarray}
where parameters $\chi'$, $\chi''$, $\bar{\chi}''$ and $\eta$ are arbitrary complex numbers, and moreover if $\chi',\chi'',\bar{\chi}''\neq0$ they are rescaling parameters. For construction of general classical $r$-matrices  (\ref{cr4}) we need to know ''overlaps'' (Schouten brackets) between the classical $r$-matrices (\ref{cr8}), (\ref{cr9}) and the Abelian $r$-matrices (\ref{cr16}). The  result of such calculation for
(\ref{cr8}) and (\ref{cr16}) is described as follows:
\begin{eqnarray}\label{cr17}
\begin{array}{rcccl}
[[a_{0}^{},\,b_{++}^{}]]\!\!&=\!\!&-2\gamma\chi' E_{+}\wedge H\wedge\bar{E}_{+}~,\qquad\;\;[[a_{0}^{},\,b_{0+}^{}]]\!\!&=\!\!&0~,
\\[7pt]
[[a_{0}^{},\,b_{+0}^{}]]\!\!&=\!\!&-2\gamma\chi'' E_{+}\wedge H\wedge\bar{H}~,\qquad\quad\;[[a_{0}^{},\,b_{00}^{}]]\!\!&=\!\!&0~,
\\[7pt]
[[a_{+}^{},\,b_{0+}^{}]]\!\!&=\!\!&-\chi\bar{\chi}''E_{+}\wedge H\wedge\bar{E}_{+}~,\qquad\quad[[a_{+}^{},\,b_{++}^{}]]\!\!&=\!\!&0~,
\\[7pt]
[[a_{+}^{},\,b_{00}^{}]]\!\!&=\!\!&-\chi\eta E_{+}\wedge H\wedge\bar{H}~,\qquad\qquad[[a_{+}^{},\,b_{+0}^{}]]\!\!&=\!\!&0~.
\end{array}
\end{eqnarray}
In order to find the overlaps between the classical $r$-matrices (\ref{cr9}) and (\ref{cr16}) we can use the involutive $\mathfrak{o}(4;\mathbb{C})$-automorphism $\varpi$ ($\varpi^2=1$): $\varpi(H)=\bar{H}$, $\varpi(E_{\pm})=\bar{E}_{\pm}$. Applying $\varpi$ to the formulas (\ref{cr17}) it is easy to find that:
\begin{eqnarray}\label{cr18}
\begin{array}{rcccl}
[[\bar{a}_{0}^{},\,b_{++}^{}]]\!\!&=\!\!&2\bar{\gamma}\chi'E_{+}\wedge\bar{E}_{+}\wedge \bar{H}~,\qquad[[\bar{a}_{0}^{},\,b_{+0}^{}]]\!\!&=\!\!&0~,
\\[7pt]
[[\bar{a}_{0}^{},\,b_{0+}^{}]]\!\!&=\!\!&2\bar{\gamma}\bar{\chi}'' H\wedge\bar{E}_{+}\wedge\bar{H}~,\qquad\quad[[\bar{a}_{0}^{},\,b_{00}^{}]]\!\!&=\!\!&0~,
\\[7pt]
[[\bar{a}_{+}^{},\,b_{+0}^{}]]\!\!&=\!\!&\bar{\chi}\chi''E_{+}\wedge\bar{E}_{+}\wedge\bar{H}~,\qquad\;[[\bar{a}_{+}^{},\,b_{++}^{}]]\!\!&=\!\!&0~,
\\[7pt]
[[\bar{a}_{+}^{},\,b_{00}^{}]]\!\!&=\!\!&\bar{\chi}\eta H\wedge\bar{E}_{+}\wedge\bar{H}~,\qquad\quad\;[[\bar{a}_{+}^{},\,b_{0+}^{}]]\!\!&=\!\!&0~.
\end{array}
\end{eqnarray}
The sums of $r$-matrices $a'$s and $b'$s in (\ref{cr17}) and (\ref{cr18}), the overlaps of which are equal to zero, 
satisfy the equation system (\ref{cr5})--(\ref{cr7}).
These $r$-matrices are given by
\begin{eqnarray}\label{cr19}
&a_{0}^{}+\,b_{00}^{},\qquad\,a_{0}^{}+\,b_{0+}^{},\qquad\,a_{+}^{}+\,b_{+0}^{},\qquad\,a_{+}^{}+\,b_{++}^{},
\\[5pt]\label{cr20}
&\bar{a}_{0}^{}+\,b_{00}^{},\qquad\,\bar{a}_{0}^{}+\,b_{+0}^{},\qquad\,\bar{a}_{+}^{}+\,b_{0+}^{},\qquad\,\bar{a}_{+}^{}+\,b_{++}^{}.
\end{eqnarray}
Comparing the solutions (\ref{cr19}) and (\ref{cr20}) we see that there are additional classical $r$-matrices
which are sums of three monomial classical $r$-matrices of type $a$, $\bar{a}$ and $b$:
\begin{eqnarray}\label{cr21}
&a_{0}^{}+\bar{a}_{0}^{}+\,\bar{b}_{00}^{},\quad\,a_{0}^{}+\bar{a}_{+}^{}+\,b_{0+}^{},\quad\,a_{+}^{}+\bar{a}_{0}^{}+\,b_{+0}^{},\quad\,a_{+}^{}+\bar{a}_{+}^{}+\,b_{++}^{},
\end{eqnarray}
{\it (ii) Case $[[b,b]]\neq0$}. If a classical $r$-matrix $r_{i}=a_i{}+\Sigma_{k}b_{k}$, where $a_i{}$ is one of (\ref{cr10}), and $b_{k}{}$ are the classical $r$-matrices (\ref{cr16}) then the compatibility
condition (\ref{cr7}) reduces to the equation for the overlaps: $\Sigma_{k<{k}'}[[b_{k},b_{k'}]]=-\Sigma_{k}[[a_{i},b_{k}]]$,
where the overlaps at the right are known already ((\ref{cr17}) and (\ref{cr18})), and we should calculate the overlaps at the left. This result is given by
\begin{eqnarray}\label{cr22}
\begin{array}{rcl}
[[b_{++}^{},\,b_{00}^{}]]\!\!&=\!\!&-\chi'\eta{E}_{+}\wedge H\wedge\bar{E}_{+}-\chi'\eta{E}_{+}\wedge\bar{E}_{+}\wedge\bar{H}~,
\\[7pt]
[[b_{++}^{},\,b_{0+}^{}]]\!\!&=\!\!&0~,\qquad\quad\,[[b_{++}^{},\,b_{+0}^{}]]\,=\,0~,\quad
\\[7pt]
[[b_{0+}^{},\,b_{00}^{}]]\!\!&=\!\!&0~,\qquad\quad\,\;\;[[b_{0+}^{},\,b_{00}^{}]]\,=\,0~,
\\[7pt]
[[b_{+0}^{},\,b_{0+}^{}]]\!\!&=\!\!&\chi''\bar{\chi}''{E}_{+}\wedge H\wedge\bar{E}_{+}+\chi''\bar{\chi}''{E}_{+}\wedge\bar{E}_{+}\wedge\bar{H}~,
\end{array}
\end{eqnarray}
Combining these formulas with ((\ref{cr17}) and (\ref{cr18})) we obtain two more classical $r$-matrices:
\begin{eqnarray}\label{cr23}
&a_{0}^{}(\gamma)-\bar{a}_{0}^{}(\gamma)-2\,b_{00}^{}(\gamma)+\,b_{++}^{}(\chi),\quad\,a_{+}^{}(\chi)+\bar{a}_{+}^{}(\chi)-\,b_{0+}^{}(\chi)+\,b_{+0}^{}(\chi).
\end{eqnarray}
Here in the $r$-matrices $a$'s, $\bar{a}$'s and $b$'s (see (\ref{cr8}), (\ref{cr9}) and (\ref{cr16})) we denote an explicit dependence on the parameter of deformation.

It should be noted that to the second solution (\ref{cr23}) we can add the term $b_{++}(\chi')$ because 
the overlaps of this term with all terms of this solution are equal to zero.
However this new solution is connected with the initial one by the automorphism which was used in the text before the formulae (\ref{cr10}). Moreover for the analysis  of the case $[[b,b]]\neq 0$ it is necessary also to take into account the basis monomials $b_{--}^{} (:=\chi{E}_{-}\wedge\bar{E}_{-})$, $b_{-+}^{}$, $b_{+-}^{}$, $b_{-0}^{}$, $b_{0-}^{}$. This consideration gives a $r$-matrix which is obtained from the first solution (\ref{cr23}) with $b_{++}^{}(\chi)$ replaced by $b_{--}^{}(\chi)$ but such new $r$-matrix is converted to the initial one by the simple  $\mathfrak{sl}(2)$-automorphisms.

Thus using $\mathfrak{sl}(2)$-grading structure of the general classical $r$-matrix ansatz (\ref{cr4}) for the complex Lie algebra $\mathfrak{o}(4;\mathbb{C})$ we found modulo $\mathfrak{sl}(2)$-automorphisms 26 classical $r$-matrices - the solutions of the classical YBE (\ref{cr1}). These solutions are given by the expressions (\ref{cr8})--(\ref{cr10}), (\ref{cr16}), (\ref{cr19})--(\ref{cr21}) and (\ref{cr23}). It is easy to see that the solutions (\ref{cr8})--(\ref{cr10}), (\ref{cr16}), (\ref{cr19})--(\ref{cr20}) can be considered as particular cases of the $r$-matrices (\ref{cr21}) if some of their deformation parameters are put equal to zero. As a result we have four three-parameter, one two-parameter and one one-parameter $r$-matrices that are described in explicit forms as follows:
\begin{eqnarray}\label{cr24}
\begin{array}{rcl}
r_{1}^{}(\gamma,\bar{\gamma},\eta)\!\!&=\!\!&\gamma\,E_{+}\wedge E_{-}+\bar{\gamma}\,\bar{E}_{+}\wedge\bar{E}_{-}+\eta\,H\wedge\bar{H},
\\[7pt]
r_{2}^{}(\gamma,\bar{\chi},\bar{\chi}'')\!\!&=\!\!&\gamma\,E_{+}\wedge E_{-}+\bar{\chi}\,\bar{E}_{+}\wedge\bar{H}+\bar{\chi}''H\wedge\bar{E}_{+},
\\[7pt]
r_{3}^{}(\bar{\gamma},\chi,\chi'')\!\!&=\!\!&\bar{\gamma}\,\bar{E}_{+}\wedge\bar{E}_{-}+\chi\,E_{+}\wedge H+\chi''{E}_{+}\wedge\bar{H},
\\[7pt]
r_{4}^{}(\chi,\bar{\chi},\chi')\!\!&=\!\!&\chi\,E_{+}\wedge H+\bar{\chi}\,\bar{E}_{+}\wedge\bar{H}+\chi'E_{+}\wedge\bar{E}_{+},
\\[7pt]
r_{5}^{}(\gamma,\chi')\!\!&=\!\!&\gamma\left(E_{+}\wedge E_{-}-\bar{E}_{+}\wedge\bar{E}_{-}-2H\wedge\bar{H}\right)+\chi'E_{+}\wedge\bar{E}_{+},
\\[7pt]
r_{6}^{}(\chi)\!\!&=\!\!&\chi(E_{+}+\bar{E}_{+})\wedge(H+\bar{H}).
\end{array}
\end{eqnarray}
Here all parameters $\gamma$, $\bar{\gamma}$, $\eta$, $\chi$, $\bar{\chi}$, $\chi'$, $\chi''$, $\bar{\chi}''$  are arbitrary complex numbers and they are independent in different $r$-matrices. It should be noted that the classical $r$-matrices $r_{2}^{}$ and $r_{3}^{}$ are connected by the involutive $\mathfrak{o}(4;\mathbb{C})$-automorphism $\varpi$: $\varpi(r_{2}^{}(\gamma,\bar{\chi},\bar{\chi}''))=r_{3}^{}(\gamma,\bar{\chi},\bar{\chi}'')$, and moreover the rest classical $r$-matrices $r_{i}^{}$ ($i=1,4,5,6$) are mapped onto themselves by the involutive automorphism $\varpi$.

Further we consider anti-Hermitian classical $r$-matrices for all $\mathfrak{o}(4;\mathbb{C})$ real forms: Euclidean, Lorentz, Kleinian and quaternionic.

\setcounter{equation}{0}
\section{Classical $r$-matrices of the $\mathfrak{o}(4;\mathbb{C})$ real forms}
It should be noted that for any classical $r$-matrix $r$, $r^{\sharp}$ ($\sharp=*,\dag,\ddag,\star$) is again a classical $r$-matrix. Moreover, the conjugations (anti-involutions) $\sharp$ retain the decomposition (\ref{cr1}), i.e. $r^{\sharp}=a^{\sharp}+\bar{a}^{\sharp}+b^{\sharp}$, where $a^{\sharp}\in A$, $\bar{a}^{\sharp}\in\bar{A}$, $b^{\sharp}\in B$ for $\sharp=*,\ddag,\star$, and $a^{\sharp}\in\bar{A}$, $\bar{a}^{\sharp}\in A$, $b^{\sharp}\in B$ for $\sharp=\dag$. All $r$-matrices (\ref{cr24}) are skew-symmetric, i.e. $r_{i}^{21}=-r_{i}^{12}$ ($i=1,\dots,6$), and further if the universal $R$-matrix $R_{r}$ of the quantum group corresponding to a real classical $r$-matrix $r$ is unitary then $r$ should be anti-Hermitian, i.e. $r^{\sharp}=-r$. In addition we shall assume that the anti-involutions $\sharp$ in (\ref{cr24}) are  lifted to the tensor product $\mathfrak{o}(4;\mathbb{C})\otimes\mathfrak{o}(4;\mathbb{C})$ by the ''flip'': ($x\otimes y)^{\sharp}=y^{\sharp}\otimes x^{\sharp}$.

Let us describe in detailed all real classical $r$-matrices for all real forms of $\mathfrak{o}(4;\mathbb{C})$.

\begin{large}\textit{1). Classical $r$-matrices of the real Euclidean algebra $\mathfrak{o}(4)$}.\end{large}
From (\ref{cr8})--(\ref{cr10}), (\ref{cr16}), (\ref{cr19})--(\ref{cr21}) and (\ref{cr23}) we see that only the $r$-matrices
\begin{eqnarray}\label{crr1}
\begin{array}{cc}
&a_{0}^{}(i\gamma),\qquad\,\bar{a}_{0}^{}(i\bar{\gamma}),\qquad\,a_{0}^{}(i\gamma)+\bar{a}_{0}^{}(i\bar{\gamma}),\qquad\,
b_{00}^{}(\eta),
\\[7pt]
&a_{0}^{}(i\gamma)+b_{00}^{}(\eta),\qquad\,\bar{a}_{0}^{}(i\bar{\gamma})+\,b_{00}^{}(\eta),\qquad\,a_{0}^{}(i\gamma)+\bar{a}_{0}^{}(i\bar{\gamma})+\,b_{00}^{}(\eta)
\end{array}
\end{eqnarray}
are anti-Hermitian with respect to the Euclidean conjugation $(^*)$ for the real parameters $\gamma,\bar{\gamma},\eta$. All these classical $r$-matrices are described by one $*$-anti-Hermitian three-parameter $r$-matrix:
\begin{eqnarray}\label{crr2}
r(\gamma,\bar{\gamma},\eta)\!\!&=\!\!&i\gamma\,E_{+}\wedge E_{-}+i\bar{\gamma}\,\bar{E}_{+}\wedge\bar{E}_{-}+\eta\,H\wedge\bar{H}
\end{eqnarray}
with the real parameters $\gamma,\bar{\gamma},\eta$.

The $r$-matrix (\ref{crr2}) can be used for the deformations of $S^3$ and $S^2\times S^{2}$ $\sigma$-models and for considering their deformed instanton solutions. We add that the compact spheres $S^{3}$ and $S^2\times S^{2}$ occur also as the internal manifolds in $D\geq 6$ string theories.

\begin{large}\textit{2). Classical $r$-matrices for the Lorentz algebra $\mathfrak{o}(3,1)$}.\end{large}
From (\ref{cr8})--(\ref{cr10}), (\ref{cr16}), (\ref{cr19})--(\ref{cr21}) and (\ref{cr23}) we see that the $r$-matrices
\begin{eqnarray}\label{crr3}
\begin{array}{cc}
&a_{0}^{}(\gamma)+\,\bar{a}_{0}^{}(\gamma^*),\quad\,b_{++}^{}(i\chi'),\quad\,b_{00}^{}(i\eta),
\\[7pt]
&a_{0}^{}(\gamma)+\,\bar{a}_{0}^{}(\gamma^*)+\,b_{00}^{}(i\eta),\quad\,a_{+}^{}(\chi)+\,\bar{a}_{+}^{}(\chi)+\,b_{++}^{}(i\chi'),
\\[7pt]
&a_{0}^{}(i\beta)-\,\bar{a}_{0}^{}(i\beta)-2\,b_{00}^{}(i\beta)+\,b_{++}^{}(i\chi'),\quad\,a_{+}^{}(\chi)+\,\bar{a}_{+}^{}(\chi)-\,b_{0+}^{}(\chi)+\,b_{+0}^{}(\chi),
\end{array}
\end{eqnarray}
are anti-Hermitian with respect to the Lorentz conjugation $(^{\dag})$ for the real parameters $\chi,\chi',\beta$ and the complex parameter $\gamma$ ($\gamma^*$ is the complex conjugation of $\gamma$). All these solutions are generated by the system of four ${\dag}$-anti-Hermitian $r$-matrices given as follows:
\begin{eqnarray}\label{crr4}
\begin{array}{rcl}
r_{1}^{}(\chi)\!\!&=\!\!&\chi(E_{+}+\bar{E}_{+})\wedge(H+\bar{H}),
\\[7pt]
r_{2}^{}(\chi,\chi')\!\!&=\!\!&\chi(E_{+}\wedge\,H+\bar{E}_{+}\wedge\bar{H})+i\chi'E_{+}\wedge\bar{E}_{+},
\\[7pt]
r_{3}^{}(\alpha,\beta,\eta)\!\!&=\!\!&(\alpha+i\beta)\,E_{+}\wedge E_{-}+(\alpha-i\beta)\bar{E}_{+}\wedge\bar{E}_{-}+i\eta\,H\wedge\bar{H},
\\[7pt]
r_{4}^{}(\beta,\chi')\!\!&=\!\!&i\beta\left(E_{+}\wedge E_{-}-\bar{E}_{+}\wedge\bar{E}_{-}-2H\wedge\bar{H}\right)+i\chi'E_{+}\wedge\bar{E}_{+},
\end{array}
\end{eqnarray}
where all arbitrary parameters $\alpha,\beta,\chi,\chi',\eta$ are real. This result coincides completely with the Zakrzewski's result \cite{Zak94} obtained by another method and presented in the $\frak{sl}(2,\mathbb{C})$-realification basis of the Lorentz algebra $\mathfrak{o}(3,1)$.

The $r$-matrices (\ref{crr4}) can be employed for the deformations of $D=4$ relativistic symmetries as well as $D=3$ de Sitter (dS) or $D=3$ hyperbolic ($H^{3}$) $\sigma$-models.

\begin{large}\textit{3). Classical $r$-matrices for the Kleinian algebra $\mathfrak{o}(2,2)$}.\end{large}
It is easy to see that all classical $r$-matrices  (\ref{cr24}) are anti-Hermitian with respect to the Kleinian conjugation $(^{\ddagger})$ for all real parameters $\gamma,\bar{\gamma},\chi,\bar{\chi},\chi',\chi'',\bar{\chi}'',\eta$. We add that some choices of these $r$-matrices used as deformations of $AdS_3$ were described by Ballesteros et all. \cite{BHMus14}--\cite{BHN15} mostly with the employments of Drinfeld double structures (see \cite{SH02}).

The classical $r$-matrices for the Kleinian algebra $\mathfrak{o}(2,2)$ are deforming $D=3$ $AdS$ geometry and can be used for the introduction of YB $\sigma$-models describing the deformations of string models with target spaces $AdS_{3}\times S^{3}$ ($D=6$; see \cite{HoTs15,LRTs14}) and $AdS_{3}\times S^{3}\times S^{3}\times T^{1}$ or $AdS_{3}\times S^{3}\times T^{4}$ ($D=10$; see \cite{Ho14}--\cite{GaGo}).

\begin{large}\textit{4). Classical $r$-matrices for the quaternionic algebra $\mathfrak{o}^{\star}(4)$}.\end{large}
It is easy to see that all anti-Hermitian classical $r$-matrices with respect to the quaternionic conjugation $(^{\star})$ are generated by the system:
\begin{eqnarray}\label{crr5}
\begin{array}{rcl}
r_{1}^{}(\gamma,\bar{\gamma},\eta)\!\!&=\!\!&\gamma\,E_{+}\wedge E_{-}+i\bar{\gamma}\,\bar{E}_{+}\wedge\bar{E}_{-}+i\eta\,H\wedge\bar{H},
\\[7pt]
r_{2}^{}(\bar{\gamma},\chi,\chi'')\!\!&=\!\!&i\bar{\gamma}\,\bar{E}_{+}\wedge\bar{E}_{-}+\chi\,E_{+}\wedge\,H+i\chi''E_{+}\wedge \bar{H},
\end{array}
\end{eqnarray}
where all parameters $\gamma$, $\bar{\gamma}$, $\eta$, $\chi$, $\chi''$ are arbitrary real numbers.

The $r$-matrices (\ref{crr5}) can be used for the construction of YB $\sigma$-models for strings with target spaces $AdS_{2}\times S^{2}$ ($D=4$; see \cite{BBHZZ99,HoTs15}),  $AdS_{2}\times S^{2}\times S^{2}$ ($D=6$; see \cite{AMPPSSTWW15}) and $AdS_{2}\times S^{2}\times T^{6}$ ($D=10$; see \cite{Ho14,AMPPSSTWW15}).

\section{Outlook}
The aim of this paper was to construct all classical $r$-matrices for the $D=4$ complex Lie algebra $\mathfrak{o}(4;\mathbb{C})$ and its real forms: Euclidean $\mathfrak{o}(4)$, Lorentz $\mathfrak{o}(3,1)$, Kleinian $\mathfrak{o}(2,2)$ and quaternionic $\mathfrak{o}^{\star}(4)$ Lie algebras. For $\mathfrak{o}(4;\mathbb{C})$ we found up to $\mathfrak{sl}(2)$-automorphisms a total list consisting of 26 classical $r$-matrices. This result was presented in the form of four three-parameter, one two-parameter and one one-parameter $r$-matrices. Employing reality conditions we obtained the classical $r$-matrices for all possible real forms of $\mathfrak{o}(4;\mathbb{C})$: compact Euclidean $\mathfrak{o}(4)$, non-compact Lorentz $\mathfrak{o}(3,1)$, non-compact Kleinian $AdS_3\cong\mathfrak{o}(2,2)$ and non-compact quaternionic $\mathfrak{o}^{\star}(4)$ Lie symmetries. For $\mathfrak{o}(3,1)$ we get known four classical $D=4$ Lorentz $r$-matrices, but for other real forms we provide new results for triangular as well as nontriangular case\footnote{Nontriangular $r$-matrices depend on the parameters $\gamma$ and $\bar{\gamma}$ which occur in the formulae (\ref{cr1}) and (\ref{cr3}).}. We can show also that for each real form the corresponding  obtained list of the classical $r$-matrices is complete up to inner automorphisms.

The next step is to obtain explicit quantizations of the given results in the spirit of our paper \cite{BLT08}. We plan to construct the complete list of classical $r$-matrices for the $D=4$ complex inhomogeneous Euclidean algebra $\mathcal{E}(4;\mathbb{C}):=\mathfrak{io}(4;\mathbb{C}):=\mathfrak{o}(4;\mathbb{C})\ltimes\mathbf{T}(4;\mathbb{C})$ (orthogonal rotations together with translations) and its real forms $\mathfrak{o}(4-k,k)\ltimes\mathbf{T}(4-k,k;\mathbb{R})$ for $k=0,1,2$.

Until present time the most complete results for $\mathfrak{o}(3,1)\ltimes\mathbf{T}(3,1)$ were obtained by Zakrzewski \cite{Zak97}, who provided almost complete list of 21 real $D=4$ Poincar\'e $r$-matrices. It should be noticed that the complete classifications of $r$-matrices for $D=3$ both Poincar\'e and Eucldean algebras have been given in \cite{Stach98}.

Recently in \cite{BLMT12,BLMT11} the present authors  complexified Zakrzewski results and then imposed $D=4$ Euclidean reality constraints. It appeared that 8 out of 21 complexified Zakrzewski $r$-matrices are consistent with the Euclidean conjugation (see (\ref{crr2})). It can be shown, however, that the complexified Zakrzewski $r$-matrices do not describe all $r$-matrices for $\mathcal{E}(4;\mathbb{C})$\footnote{In particular one can argue that the list of the real $r$-matrices for $\mathfrak{o}(2,2)\ltimes\mathbf{T}(2,2)$ is longer then the Zakrzewski list for the $D=4$ Poincar\'e algebra.}.

We add that in \cite{BLMT12,BLMT11} we considered also the $N=1$ superextension of Poincar\'e and Euclidean classical $r$-matrices. Recently we derived in analogous way also new class of $N~=~2$ Poincar\'e and Euclidean supersymmetric $r$-matrices (see \cite{BLT15}). We hope, however, that our constructive method can be used to provide a complete list of the $r$-matrices for the complex inhomogeneous algebra $\mathcal{E}(4;\mathbb{C})$, for its real forms, and then it can be applied to $N$-extended Euclidean superalgebras $\mathcal{E}(4|N;\mathbb{C})$ (in particular for physically important cases $N=1,2,4$ containing for $N=2,4$ the central charges). The final aim is to classify the $N$-extended supersymmetric $r$-matrices for all corresponding supersymmetrized real forms. We add that such super extension is necessary if we consider the deformations of $D=10$ critical superstring, reduced further to $D=4$.

\subsection*{Acknowledgments}
The authors would like to thank P.~Aschieri,  S.~van Tongeren and K.~Yoshida for valuable comments. The work has been supported by Polish National Science Center (NCN), project 2014/13/B/ST2/04043 (A.B. and J.L.) and by COST (European Cooperation in Science and Technology) Action MP1405 QSPACE (A.B. and J.L.). V.N.T. is supported by Russian RFBR grant No.14-01-00474-a.
The first author acknowledges hospitality from the Department of Sciences and Technological Innovation, University of Eastern Piedmont during the preparation of the manuscript.


\begin{thebibliography}{99}

\bibitem{Maj88} S.~Majid, J. Class. Quant. Grav. \textbf{5}, 1587 (1988).

\bibitem{DFR95} S.~Doplicher, K.~Fredenhagen, J.E.~Roberts, Commun. Math. Phys. \textbf{172}, 187 (1995); arXiv:hep-th/0303037.

\bibitem{Gar95} J.L. Garay, Int. Jour. Math. Phys. \textbf{A10}, 145 (1995); arXiv:gr-qc/9403008.
\bibitem{BD82} A.A. Belavin, V.G. Drinfeld, Funct. Anal. Appl. \textbf{16}, 1 (1982).

\bibitem{Sem-T-Sh83} M.A.~Semenov-Tian-Shansky, Funct. Anal. Appl. \textbf{17}, 289 (1983).

\bibitem{ChPr94}V.~Chari, A.~Pressley, ``A Guide to Quantum Groups'', Cambridge Univ. Press, 1994.

\bibitem{Maj95} S.~Majid, ``Foundations of  Quantum Groups'', Cambridge Univ. Press, 1995.

\bibitem{HZ09} C.~Hull, B.~Zwiebach, JHEP \textbf{0909}, 099 (2009); arXiv:0904.4664[hep-th].

\bibitem{Ren14} F.~Rennecke, JHEP \textbf{1410}, 069 (2015); arXiv:1404.0912[hep-th].

\bibitem{Klim02} C.~Klimcik, JHEP \textbf{0212}, 051 (2002); arXiv:0210.095[hep-th].

\bibitem{Klim09} C.~Klimcik, J. Math. Phys. \textbf{50}, 043508 (2009); arXiv:0802.3518[hep-th].

\bibitem{Vic15}B.~Vicedo, J. Phys. \textbf{A48}, 355203 (2015); arXiv:1504.06303[hep-th].

\bibitem{KMY14} T.~Kawaguchi, T.~Matsumoto, K.~Yoshida, JHEP \textbf{1404}, 153 (2014); arXiv:1401.4855[hep-th]; see also JHEP \textbf{1406}, 146 (2014); arXiv:1402.6147[hep-th].

\bibitem{MY14} T.~Matsumoto, K.~Yoshida, J. Phys.: Conf. Ser. \textbf{56}, 012020 (2015); arXiv:1410.0575[hep-th].

\bibitem{MY15} T.~Matsumoto, K.~Yoshida, Nucl. Phys. \textbf{B893}, 287 (2015); arXiv:1501.03665.

\bibitem{vanTong15} S.J.~van Tongeren, JHEP \textbf{1506}, 048 (2015); arXiv:1504.05516[hep-th]; see also arXiv:1506.01023[hep-th].

\bibitem{PolTong15} A.~Pachol, S.J.~van Tongeren,  arXiv:1510.02389[hep-th].

\bibitem{BKLSY15} A.~Borowiec, H.~Kyono, J.~Lukierski, J.~Sakamoto, K.~Yoshida, arXiv:1510.03083[hep-th].

\bibitem{HI99} A.~Hashimoto, N.~Itzaki, Phys. Lett. \textbf{B465}, 142 (1999); arXiv:hep-th/9907166.

\bibitem{MR99} J.M.~Maldacena, J.G.~Russo, JHEP \textbf{9909}, 025 (1999); arXiv:hep-th/9908134.

\bibitem{MYosh14} T.~Matsumoto, K.~Yoshida, JHEP \textbf{1406}, 163 (2014); arXiv:1404.3657[hep-th].

\bibitem{Zak94} S.~Zakrzewski, Lett. Math. Phys. \textbf{32}, 11 (1994).

\bibitem{Tol07} V.N.~Tolstoy, Bulg. J. Phys.  \textbf{35}, 441 (2008) (Conference: C07-06-18.13 Proceedings); arXiv:0712.3962.

\bibitem{Gilm74} R.~Gilmore, ``Lie Groups, Lie Algebras and Some of Its Applications'',  New York 1974.

\bibitem{Gant39} F.R.~Gantmacher, Sbornik: Mathematics, \textbf{5}, 217 (1939).

\bibitem{BarutRacz77} A.O.~Barut, R.~Raczka, ''Theory of group representations and application'', PWN, Warsaw 1977.

\bibitem{D-VL90} M.~Dubois-Violette, G.~Launer, Phys. Lett. \textbf{B275}, 175 (1990).

\bibitem{BHMus14} A.~Ballesteros, F.J.~Herranz, F.~Musso, J. Phys.: Conf. Ser. \textbf{532}, 012002 (2014); arXiv:1302.0684[hep-th].

\bibitem{BHMeas13} A.~Ballesteros, F.J.~Herranz, C..~Meusburger, Class. Quant. Grav. \textbf{30}, 155012 (2014); arXiv:1303.3080[hep-th].

\bibitem{BHN15} A.~Ballesteros, F.J.~Herranz, P.~Naranjo, Phys. Lett. \textbf{B746}, 37 (2015); arXiv:1502.07518[gr-qc].

\bibitem{SH02} L.~Snobl, L.~Hlavaty,  Internat. J. Modern Phys. \textbf{A17} (28), 4043 (2002).

\bibitem{HoTs15} B.~Hoare, A.A.~Tsetlin, JHEP \textbf{1510}, 060 (2015); arXiv:1508.01150[hep-th].

\bibitem{LRTs14} O.~Lunin, R.~Roiban,  A.A.~Tsetlin,  Nucl. Phys. \textbf{B891}, 106 (2015); arXiv:1411.1066[hep-th].

\bibitem{Ho14} B.~Hoare,  Nucl. Phys. \textbf{B891}, 259 (2015); arXiv:1411.1266[hep-th].

\bibitem{AMPPSSTWW15} A.~Abbort, J.~Murugan, S.~Penatiu, A.~Pittelli, D.~Sorokin, P.~Sundin, J.~Tarrant,  M.~Wolf, L.~Wulf,  arXiv:1509.07678[hep-th].

\bibitem{GaGo} A.R.~Gaberdiel, R.~Gopakumar,   arXiv:1512.07237[hep-th].

\bibitem{BBHZZ99}N.~Berkovits, M.~Bershadsky, T.~Hauer, S.~Zhokov, B.~Zwiebach,  Nucl. Phys. \textbf{B567},  61 (2000); arXiv:hep-th/9907200.

\bibitem{BLT08} A.~Borowiec, J.~Lukierski, V.N.~Tolstoy, Eur. Phys. J. \textbf{C57}, 601 (2008); arXiv:0804.3305[hep-th].

\bibitem{Zak97} S.~Zakrzewski, Commun. Math. Phys. \textbf{178}, 285 (1997); arXiv:q-alg/9602001.

\bibitem{Stach98} P. Stachura,
J. Phys. \textbf{A 31}, 4555 (1998).

\bibitem{BLMT12} A.~Borowiec, J.~Lukierski, M. Mozrzymas, V.N.~Tolstoy, JHEP \textbf{1206}, 154 (2012); arXiv:1112.1936[hep-th].

\bibitem{BLMT11} A.~Borowiec, J.~Lukierski, M. Mozrzymas, V.N.~Tolstoy, Proc. XXIX Jnt. Coll. on Group-Theoretical Methods in Physics, Tianjin, August 2012,  ed. Cheng-Ming Bai et all., World Scientific, Singapore, p.443 (2013); arXiv:1211.4546[hep-th].

\bibitem{BLT15} A.~Borowiec, J.~Lukierski, V.N.~Tolstoy,  arXiv:1510.09125[hep-th].

\end{thebibliography}
\end{document}